\documentclass[aip,jcp,amsmath,amssymb,reprint,floatfix]{revtex4-1}

\usepackage{amssymb}
\usepackage{mathptmx}
\usepackage{color}

\usepackage{graphicx}
\usepackage{dcolumn}
\usepackage{bm}

\usepackage[utf8]{inputenc}

\usepackage[T1]{fontenc}
\usepackage{mathptmx}
\usepackage{etoolbox}
\usepackage{multirow}
\usepackage[parfill]{parskip}
\usepackage{float}
\usepackage[colorlinks,allcolors=blue]{hyperref}
\makeatletter
\def\@email#1#2{%
 \endgroup
 \patchcmd{\titleblock@produce}
  {\frontmatter@RRAPformat}
  {\frontmatter@RRAPformat{\produce@RRAP{*#1\href{mailto:#2}{#2}}}\frontmatter@RRAPformat}
  {}{}
}
\makeatother

\draft

\newcommand{\duo}{\textsc{Duo}}
\newcommand{\Duo}{\textsc{Duo}}

\newcommand{\Python}{\textsc{Python}}
\newcommand{\binslt}{\textsc{binSLT}}

\newcommand{\MolPro}{\textsc{MolPro}}

\newcommand{\cm}{cm$^{-1}$}
\newcommand{\re}{$r_\textrm{e}$~}
\newcommand{\rc}{$r_\textrm{c}$~}
\newcommand{\Te}{$T_\textrm{e}$~}
\newcommand{\Ae}{$A_\textrm{e}$~}

\newcommand{\ai}{\textit{ab initio}~}
\newcommand{\AI}{\textit{Ab initio}~}

\newcommand{\X}{X~$^2\Pi$}
\newcommand{\Xx}{X~$^2\Pi_x$}
\newcommand{\Xy}{X~$^2\Pi_y$}

\newcommand{\A}{A~$^2\Sigma^+$}
\newcommand{\B}{B~$^2\Sigma^+$}
\newcommand{\C}{C~$^2\Sigma^+$}

\newcommand{\IS}{1~$^2\Sigma^-$}
\newcommand{\IQS}{1~$^4\Sigma^-$}
\newcommand{\IQPi}{1~$^4\Pi$}
\newcommand{\IQPix}{1~$^4\Pi_x$}
\newcommand{\IQPiy}{1~$^4\Pi_y$}
\newcommand{\AllI}{1~$(^2\Sigma^-,^4\Sigma^-,^4\Pi)$}

\newcommand{\HSO}{$|\hat{\textrm{H}}_{\textrm{SO}}|$}

\newcommand{\Lx}{$|\hat{\textrm{L}}_{\textrm{x}}|$}

\newcommand{\XSOA}      {$\langle$\X\HSO\A$\rangle$}
\newcommand{\XLxA}      {$\langle$\X\Lx\A$\rangle$}
\newcommand{\XSOX}      {$\langle$\X\HSO\X$\rangle$}
\newcommand{\ASOIS}     {$\langle$\A\HSO\IS$\rangle$}
\newcommand{\ASOIQS}    {$\langle$\A\HSO\IQS$\rangle$}
\newcommand{\ASOIQPi}   {$\langle$\A\HSO\IQPi$\rangle$}

\newcommand{\quantumnumbers}{$\{J,v,\tau\}$}

\newcommand{\AState}    {$|\textrm{A} ^2\Sigma^+, Q\rangle$}

\begin{document}

\title{Predissociation dynamics of the hydroxyl radical (OH) based on a five-state spectroscopic model} 

\author{G. B. Mitev}
\email{georgi.mitev.16@ucl.ac.uk}
\author{Jonathan Tennyson}
\email{j.tennyson@ucl.ac.uk}
\author{Sergei N. Yurchenko}
\affiliation{Department of Physics and Astronomy, University College London, Gower St, London WC1E 6BT, UK}
\date {\today}
\begin{abstract}
Multi-reference configuration interaction (MRCI) potential energy curves (PECs) and spin-orbit couplings for the  
\X, \A,  1 $^2 \Sigma^-$, 1 $^4 \Sigma^-$, and 1 $^4 \Pi$  states of OH are computed and refined against empirical energy levels and transitions to produce a spectroscopic model. Predissociation lifetimes are determined by discretising continuum states in the variational method nuclear motion calculation by restricting the calculation to finite range of internuclear separations. Varying this range give a series of avoided crossings between quasi-bound states associated with the A $^2 \Sigma^+$ and continuum states, from which predissociation lifetimes are extracted.  424 quasi-bound \A\  state rovibronic energy levels are analysed and 374 predissociation lifetimes are produced, offering good coverage of the predissociation region.  Agreement with measured lifetimes is satisfactory and a majority of computed results were within experimental uncertainty. A previously unreported \A\  state predissociation channel which goes via the \X\ is identified in the calculations.
A python package, \binslt, is produced to calculate predissociation lifetimes, associated line broadening parameters, and uncertainties from \Duo\ *.states files is made available.
The PECs and other curves from this work will be used to produce a rovibronic ExoMol linelist and temperature-dependent photodissociation cross sections for the hydroxyl radical.
\end{abstract}
\maketitle


\section{Introduction}
The hydroxyl radical OH is of significance in a diverse set of physical systems and as such has been extensively studied. OH is of high importance due to its presence in combustion, atmospheric and interstellar chemistry, and as a key constituent of the Earth's atmosphere \cite{20SuZhQi.OH,03ZhYuZh.OH, 08RaRoWu.OH, 84NeLe.OH, 01Joens.OH, 95PrWe.OH,16BrBeWe.OH}. Furthermore, OH has been detected recently in the atmosphere of Ultra-hot Jupiters WASP-76b and WASP-33b \cite{21LaSaMo.OH,23WrNuBr.OH}. 
The many high resolution spectroscopy studies on OH have recently been comprehensively reviewed by Furtenbacher {\it et al.} \cite{jt868}
as part of their MARVEL (measured active rotation energy level) study.
The \ai electronic structure and predissociation dynamincs of OH have been of interest for many years. Much of the early theoretical work was done by Langhoff, van Dishoeck, Dalgarno, Bauschlicher and Wetmore \cite{83DiDa.OH, 82LaDiWe.OH, 83DiLaDa.OH, 84DiDa.OH, 87BaLa.adhoc}, these works have seen extensive use in other theoretical studies \cite{92Yarkony.OH, 87LeFr.OH, 94KaSa.OH, 95Le1.OH, 95Le2.OH, 95Le3.OH, 95Le4.OH, 96Le.OH, 84DiHeAl.OH}. Further \ai studies have been completed more recently with more computational power and larger basis sets \cite{05LoGr.OH, 14SrSaxx.OH, 14QiZhxx.OH, 92Yarkony.OH, 99PaYa.OH}. We particularly highlight the work of \citet{05LoGr.OH} which provided a starting point this study.

Despite the need for high-accuracy potential energy curves and coupling curves, the angular momentum coupling curves do not seem to have been reported. Angular momentum coupling corresponds to the $\Lambda$-doubling in the ground \X\  electronic state energy levels \cite{jt632} and are required for accurate modelling of the $e/f$ parity splittings in both \X\  and \A\  states. This splitting has been reported to be anomalously high at ultra-cold temperatures \cite{13RaLiDo.OH}, hence, the angular momentum coupling between the \X\  and first electronic excited state, \A\  should also be of interest to ultra-cold physics experiments. 
The effect of predissociation of the A $^2 \Sigma^+$ energy levels caused by a spin-orbit interaction with repulsive (unbound) electronic states 1,$^2\Sigma^-$, 1,$^4\Sigma^-$ and 1,$^4 \Pi$ is investigated. Predissociation is one of the main sources of line broadening in the A--X rovibronic transitions, and has been extensively studied both experimentally and theoretically, for which lifetimes, line positions, line widths, rates, and branching ratios have been reported \cite{05DePoDe.OH, 78BrErLy.OH, 91GrFa.OH,92HeCrJe.OH,97SpMeMe.OH, 21SuZhZh.OH, 11LiZh.OH, 94KaSa.OH, 95Le1.OH, 95Le2.OH, 95Le3.OH, 95Le4.OH, 96Le.OH, 87LeFr.OH, 80SiBaLe.OH,84DiHeAl.OH}. A summary of available lifetimes data can be found is given below in Table~\ref{tab:prediss_summary}.
In the next section, details of the electronic structure calculations and spectroscopic model refinement procedure are presented. The method
used to compute predissociation lifetime is based on use of variational bound-state nuclear-motion program \duo.\cite{jt609} Section 3 discusses how \duo\ is used to study predissocation; a fuller discussion of this method will be presented in a paper\cite{jtpred} henceforth referred to as Paper I.
Section \ref{sec:results} presents the final spectroscopic model and compares our calculated lifetimes with their literature counterparts. A summary of findings and proposed future work are presented in section \ref{sec:conclusion}.


\section{Methods: \AI}
\label{sec:meth_ai}
The aim of this study is to produce a complete set of predissociation lifetimes for the \A\ state in the OH radical. An accurate spectroscopic model is needed to perform the lifetime calculations. To produce this model, a high level of theory was employed to calculate \ai PECs and coupling curves for the \X, \A, and the dissociative \AllI\ states, see Fig.~\ref{fig:PECs}. These are  refined against experimental values in sec. \ref{subsec:fitting}.
\subsection{\AI electronic structure calculations}
\label{subsec:molpro}
The initial potential energy curves (PECs), spin-orbit coupling curves (SOCs) and angular momentum coupling curves (AMCs) were computed using the \MolPro~quantum chemistry program \cite{molpro.method, MOLPRO, MOLPRO2020}. 
Following \citet{05LoGr.OH}, optimal molecular orbitals were computed using carefully selected combinations of state-averaged complete active space self-consistent field (SA-CASSCF) \cite{85WeKn.adhoc} calculations: details of which are given in Table \ref{tab:molpro}.
These orbitals provide the input to multi-reference configuration interaction (MRCI) calculations which included a
Davidson correction \cite{74LaDa.adhoc} to the energies. Final results were computed using an aug-cc-pV6Z basis set.  The calculations were performed over internuclear distances ranging from 1 to 10 $\textrm{a}_0$ with a greater density of points around the equillibrium bond length. We ensured that no $^2\Delta$ states were obfuscating the presence of the desired $^2 \Sigma^{\pm}$ states by calculating PECs for both $A_1$ and $A_2$ irreducible  representations of $C_{2v}$  and selecting the appropriate symmetries. 
There are two linearly independent spin components in the $^4\Pi$ state giving rise to the same value of $|\Omega| = \frac12$: $|1'~^4\Pi\rangle \rightarrow \Lambda = -1, \Sigma = \frac32$ and $|1~^4\Pi\rangle \rightarrow \Lambda = 1, \Sigma = -\frac12$. 

\begin{figure}[H]
\includegraphics[width=0.5\textwidth]{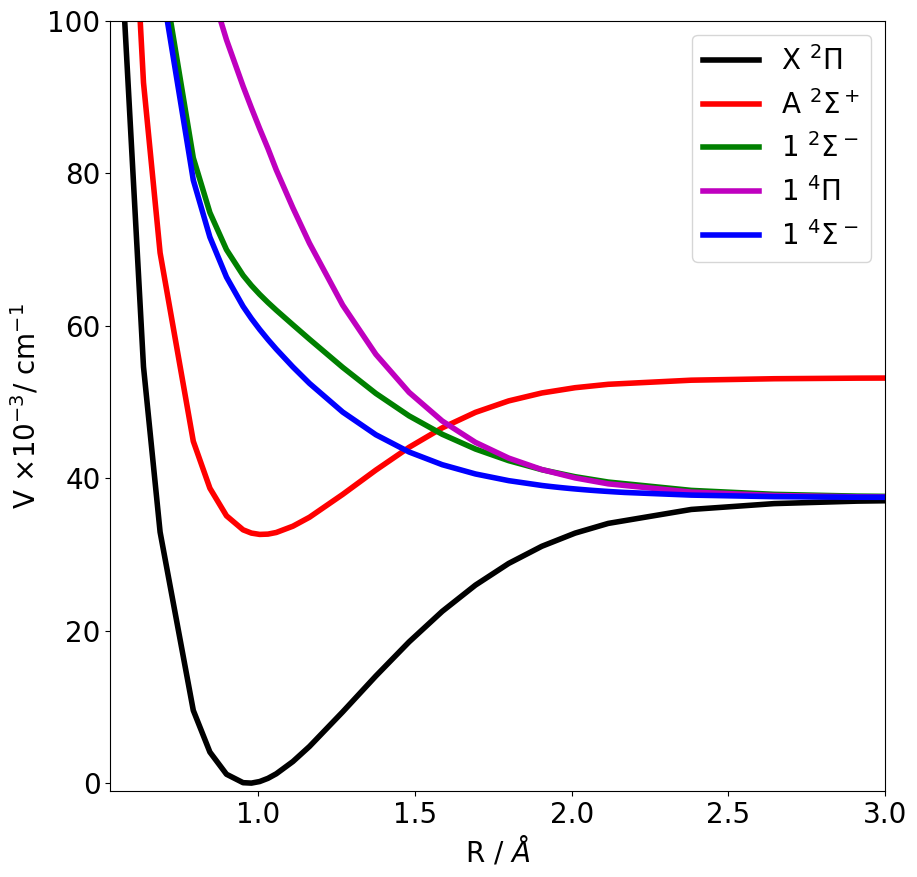}
\caption{\label{fig:PECs} \AI potential energy curves for OH as computed in \MolPro.}
\end{figure}

\begin{figure}[H]
\includegraphics[width=0.5\textwidth]{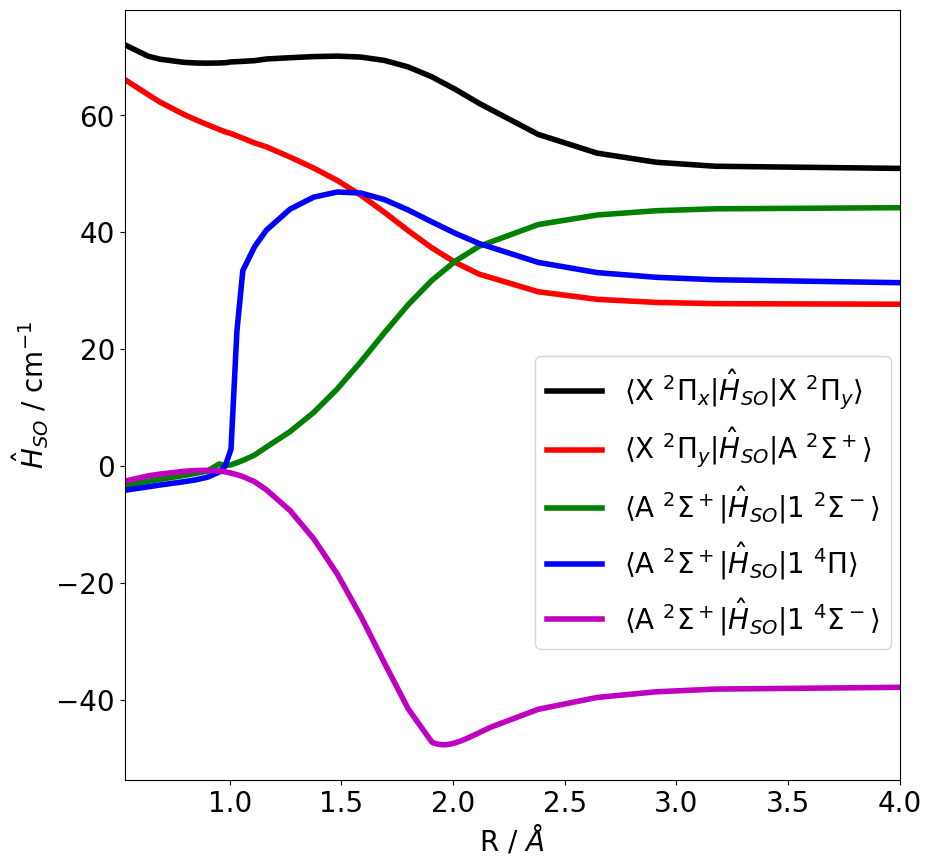}
\caption{\label{fig:SOCs} Diagonal and off-diagonal \ai spin-orbit coupling curves of OH as computed in \MolPro.  The curves here also have an associated factor of $\pm i$.}
\end{figure}

\begin{figure}[H]
\includegraphics[width=0.5\textwidth]{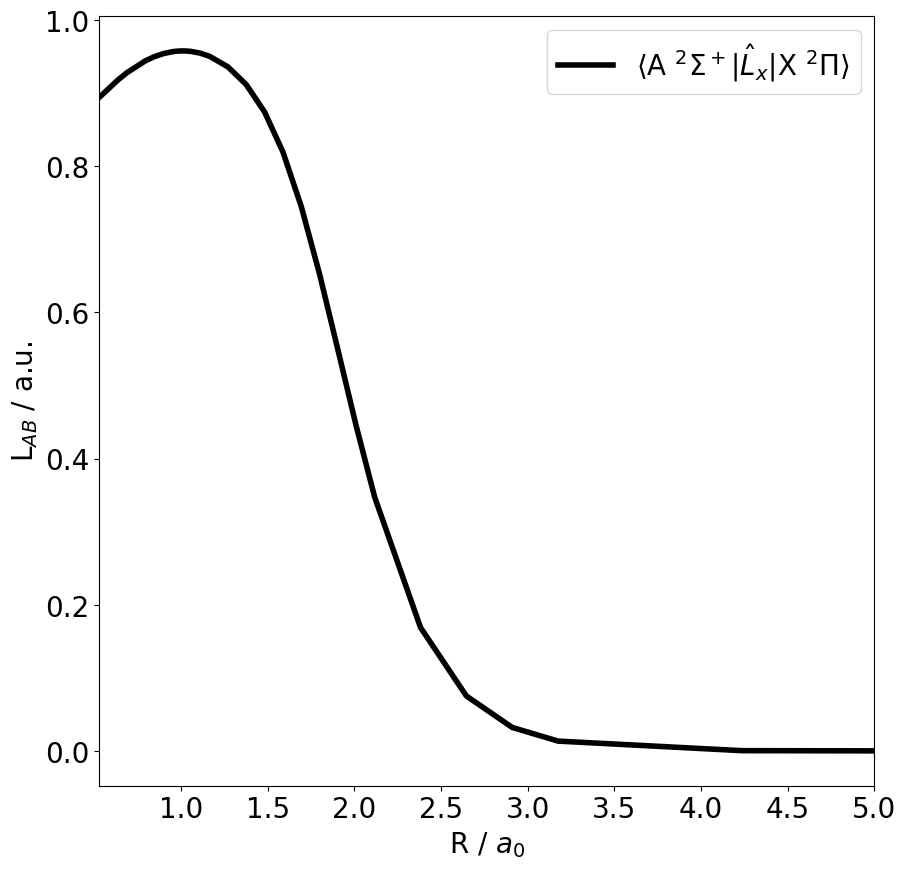}
\caption{\label{fig:L} \AI L-uncoupling matrix elements for OH as computed in \MolPro.  In the cartesian coordinate system, there is a factor of $\pm i$.}
\end{figure}

The spin-orbit coupling between the \A\ and \IQPi\ states have the relationship, 
\begin{equation}
\langle 1~^4\Pi|\hat{\textrm{H}}_\textrm{SO}|\textrm{A}~^2\Sigma^+\rangle = -\sqrt{3}\langle 1'~^4\Pi|\hat{\textrm{H}}_\textrm{SO}|\textrm{A}~^2\Sigma^+\rangle
\end{equation}
derived from the application of the Wigner-Eckart theorem \cite{99PaYa.OH, 92Yarkony.OH}. In \Duo, for spin-orbit matrix elements it is sufficient to specify only one of these combinations with the correct $\Lambda$ and $\Sigma$, the other is generated using the Wigner-Eckart theorem. 
Here we select 
$1~^4\Pi$, where $\Sigma = -\frac12$, $\Lambda = 1$ as is done in \citet{99PaYa.OH}. One should note it does not matter which coupling is chosen, as long as the correct $\Lambda,~\Sigma$ combination is given in Duo. 
\MolPro~produces coupling curves and dipoles with an arbitrary phase factor of $\pm 1$ or $\pm i$ 
which is not guaranteed consistent between geometries. This uncertainty in phase leads to discontinuities in the curves and hence requires post-processing. Within an MRCI calculation informed by a set of SA-CASSCF orbitals, the phase may not be consistent between geometries, however it is consistent over all curves computed with those orbitals for a given geometry; any discontinuities will appear in the same places for all curves \cite{jt589}. The coupling curves in this study, however, were not all computed with one set of orbitals and instead were split up as shown in the SA-CASSCF column in Table \ref{tab:molpro}. Inter-SA-CASSCF phase consistency was ensured by first smoothing all curves with an arbitrary global phase and comparing to a set of reference curves. These were produced by performing an aug-cc-pVTZ calculation with all states present in the SA-CASSCF, ensuring phase consistency between all relevant curves. Visual representations of these curves can be seen in Figs. \ref{fig:PECs}, \ref{fig:SOCs}, and \ref{fig:L}
\begin{table}[H]
\caption{\label{tab:molpro}Summary of \ai electronic structure calculation details. All calculations use aug-cc-pV6Z basis set with the first $\sigma$ orbital closed. Orbitals correspond to the point group symmetry $C_{2v}$.}
\begin{ruledtabular}
\begin{tabular}{llr}
Curve     & SA-CASSCF          & Space              \\
\hline
PECs            &&                                  \\
\hline          
\X        & \Xx,\Xy            & $5\sigma2\pi$      \\
\A, \IS   & \A, \IS            & $5\sigma2\pi$      \\
\IQS      & \A, \IS, \IQS      & $5\sigma2\pi$      \\
\IQPi     & \IQPix,\IQPiy, \A  & $6\sigma2\pi$      \\
\hline
Coupling Curves &&                                  \\
\hline          
\XSOA     & \Xx,\Xy, \A        & $5\sigma2\pi$      \\
\XLxA     & \Xx,\Xy, \A        & $5\sigma2\pi$      \\
\XSOX     & \Xx,\Xy            & $5\sigma2\pi$      \\
\ASOIS    & \A, \IS            & $5\sigma2\pi$      \\
\ASOIQS   & \A, \IS, \IQS      & $5\sigma2\pi$      \\
\ASOIQPi  & \IQPix,\IQPiy, \A  & $6\sigma2\pi$      \\
\end{tabular}
\end{ruledtabular}
\end{table}


\section{METHODS: Hydroxyl spectroscopic model}
\label{fitting}
\subsection{Global parameters}
\label{subsec:global}
The curves produced above provide input to  the nuclear motion code \Duo  \cite{Duo}, an open-source Fortran 2009 program which provides variational solutions to the coupled rovibronic Schr\"{o}dinger equations for a general open-shell diatomic molecule. Rovibronic energy levels and transitions are evaluated for the mutually perturbed \X\  and \A\  states. However, the \ai curves are insufficient for accurate reproduction and extrapolation of energy levels for this system despite the high level of \ai\ 
theory employed.

A number of global parameters need to be established before running a \Duo~calculation, they are as follows: The size of vibrational basis sets ($v_{\rm max}$) for each electronic state, the start ($r_{\rm min}$) and end ($r_{\rm c}$) of the calculation box, the number of grid points, and the maximum value for $J$. These considerations are made to ensure converged solutions for the energy levels. The start and end of the box was set to 0.53 \AA\ and 8 \AA~respectively with a grid size of 801 points. 
The density of rovibronic states increases with energy. States near the top end of the available vibrational basis are unstable as they are asymmetrically perturbed by local rovibronic states due to lack of states at $v>v_{\rm max}$ which means that an insufficiently large $v_{\rm max}$ may lead to  bound state solutions that may not be stable upon variation of the box size. $v_{\rm max}$ was selected for each electronic state iteratively. By extracting energy eigenvalues within the relevant fitting region for increasing $v_{\rm max}$, we can set $v_{\rm max}$ such that the energy levels are converged. This was found to occur when $v_{\rm max}^X \geq 110$ and $v_{\rm max}^A \geq 90$. The size of the vibrational basis sets were hence set to 110 and 90 for the \X\  and \A\  states respectively. The maximum value for $J$, $J_{\textrm{max}}$, was set to 50.5.  
\subsection{Fitting}
\label{subsec:fitting}
The PECs and coupling curves were refined by making constrained adjustments to parameters which describe them with respect to a set of observed data, which can be either empirical energy levels or transitions. These data are supplied by the recent MARVEL data set \cite{jt868}. Energy level coverage of this dataset is summarized in table \ref{tab:marvel_coverage}. Curve refinement is performed in \Duo~by least squares fitting to this data set. 
\begin{table}
\caption{\label{tab:marvel_coverage} Coverage of MARVEL data set \cite{jt868} energy levels for the \X\  and \A\  state of OH which have been used to refine \ai PECs, SOCs and coupling constants for the production of a spectroscopic model.}
\begin{ruledtabular}
\begin{tabular}{rrrr}
\multicolumn{1}{c}{v} & \multicolumn{1}{c}{$J_{\textrm{min}}$} & \multicolumn{1}{c}{$J_{\textrm{max}}$} & \multicolumn{1}{c}{Number of Levels} \\ 
\multicolumn{4}{c}{\X}                                                                                           \\ \hline
0                     & 0.5                     & 31.5                    & 126                                  \\
1                     & 0.5                     & 31.5                    & 123                                  \\
2                     & 0.5                     & 31.5                    & 123                                  \\
3                     & 0.5                     & 31.5                    & 122                                  \\
4                     & 0.5                     & 19.5                    & 77                                   \\
5                     & 0.5                     & 19.5                    & 76                                   \\
6                     & 0.5                     & 18.5                    & 74                                   \\
7                     & 0.5                     & 18.5                    & 74                                   \\
8                     & 0.5                     & 18.5                    & 74                                   \\
9                     & 0.5                     & 18.5                    & 74                                   \\
10                    & 0.5                     & 11.5                    & 42                                   \\
11                    & 0.5                     & 8.5                     & 23                                   \\
12                    & 0.5                     & 7.5                     & 28                                   \\
13                    & 0.5                     & 7.5                     & 27                                   \\ 
Total                 & 0.5                     & 31.5                    & 1063                                 \\ \hline
\multicolumn{4}{c}{\A}                                                                                           \\ \hline
0                     & 0.5                     & 31.5                    & 64                                   \\
1                     & 0.5                     & 28.5                    & 58                                   \\
2                     & 0.5                     & 19.5                    & 40                                   \\
3                     & 0.5                     & 26.5                    & 54                                   \\
4                     & 0.5                     & 19.5                    & 40                                   \\
5                     & 0.5                     & 7.5                     & 15                                   \\
6                     & 0.5                     & 8.5                     & 17                                   \\
7                     & 0.5                     & 7.5                     & 16                                   \\
8                     & 0.5                     & 9.5                     & 19                                   \\
9                     & 0.5                     & 8.5                     & 17                                   \\ 
Total                 & 0.5                     & 31.5                    & 340                                  \\ \hline
Grand Total           &                         &                         & 1403                                
\end{tabular}
\end{ruledtabular}
\end{table}
\Duo~provides three types of fitting one can use; variation of parameters that describe a curve (sec. \ref{subsubsec:prefit}), variation of parameters that describe a morphing function (\textit{morphing}, see sec. \ref{subsubsec:morphing}), and direct variation of individual grid points. In each case the optimization is based on a non-linear conjugate gradient method \cite{Duo} with respect to an empirical data set. 
\subsubsection{\textit{Parameter variation fitting} -- pre-fitting}
\label{subsubsec:prefit}
During parameter variation fitting, the curves must have some parametric form, and some initial set of parameters must be determined. For PECs, the form used in this study is the Extended Morse Oscillator (EMO)\cite{dPotFit,Duo}, given by
\begin{equation}\label{eq:EMO}
    V_{\rm EMO}(r) = T_{\rm e} + (A_{\rm e} - T_{\rm e}) \left( 1 - e^{-\beta(r)(r-r_{\rm e})} \right) ^2
\end{equation}    
where \Te is the potential minimum, $ D_{\rm e} = A_{\rm e} - T_{\rm e}$ is the dissociation energy relative to \Te, $A_{\rm e}$ is the dissociation asymptote, \re is the equilibrium bond length, and $\beta(r)$ is a distance dependent exponent coefficient defined by the expansion term with coefficients $\beta_i$ with respect to the reduced coordinate $\xi_p$, first introduced by \citet{84SuRaBo.method} add are given by
\begin{eqnarray}\label{eq:beta}
     \beta(r) = \sum^{N}_{i=0}\beta_i \xi_p(r)^i;\quad
         N =
    \begin{cases}
    N_R & \textrm{for} ~~ r > r_{\rm e} \\
    N_L & \textrm{for} ~~ r \leq r_{\rm e} ,
    \end{cases}
\end{eqnarray}   
\begin{eqnarray}
\label{eq:surkus}
    \xi_p(r) = \frac{r^{p}-r_{\rm e}^{p}}{r^{p}+r_{\rm e}^{p}};\quad
         p =
    \begin{cases}
    p_R & \textrm{for} ~~ r > r_{\rm e} \\
    p_L & \textrm{for} ~~ r \leq r_{\rm e} .
    \end{cases}
\end{eqnarray}
This parametric form is flexible in that one can specify the behaviour around the equilibrium internuclear distance piece-wise using the parameters $N_L$, $N_R$, $p_L$, and $p_R$, where $N_L \leq N_R$. The value of $N_R$ hence defines the order of the expansion. This means one can establish a family of potential energy curves where the fundamental shape is determined by the set of parameters, $S$, where
\begin{eqnarray}\label{eq:shape}
    S = \{p_L,~p_R,~N_L,~N_R\}; ~~ \{p_L,~p_R,~N_L,~N_R\} \in \mathbb{Z},~N_R\geq N_L.
\end{eqnarray}   
For a given $S$, one can then find $P$ which refines the shape to return appropriate energy levels, such that: 
\begin{eqnarray}\label{eq:setP}
    P = \{\beta_i,~r_{\rm e},~T_{\rm e},~D_{\rm e}\}; ~~i \leq N_R.
\end{eqnarray}   
When fitting a PEC in \Duo, $S$ is fixed, and hence should be carefully chosen before optimizing $P$.  Initial parameter selection has been performed by extending the method of \citet{21MiTaTe.NaO}. Given that the elements of $S$ are integers, one cannot use a standard optimization algorithm to optimize the values, and instead must solve the problem iteratively. The optimization of $S$ is therefore initiated by establishing the set,
\begin{eqnarray}
    \sigma_{S} =&  \{S_0,...,S_n\}  \\
    \equiv& \{\{p_{L_0},p_{R_0},N_{L_{0}}, N_{R_{0}}\},...,\{p_{L_n},p_{R_n},N_{L_{n}}, N_{R_{n}}\}\}
\end{eqnarray}
which contains $n$ elements of the set, $S$, as in Eq.~(\ref{eq:shape}) such that,
\begin{eqnarray}
    S_i \neq S_j ~\forall~ i,j ;~~ i \neq j.
\end{eqnarray}
This is the set of values of $p_L$, $p_R$, $N_L$, and $N_R$ over which we would like to test. The test consists of, for each $S_i$, finding the conjugate $P_i$ by least-squares fitting against the \ai grid points in \Python\ (code available on \href{https://github.com/exomol}{ExoMol GitHub}). In each case we find the reduced $\chi^2$ test statistic, $\chi^2_{\nu}$ and search for the $S$, $P$ which return the lowest value. These parameters are then used as a starting point for fitting the PECs against experimental data in \Duo. 
See table \ref{tab:sigmatot} for the $\sigma_i$ parameters for the \X\  and \A\  states in this study. 
\begin{table}[H]
\caption{\label{tab:sigmatot} Parameters, $\sigma_i$ which best describe \ai PECs of OH as calculated in sec \ref{subsec:molpro}. Units in table footnote.}
\begin{ruledtabular}
\begin{tabular}{lrlr}
\multicolumn{2}{c}{\X} & \multicolumn{2}{c}{\A} \\
\hline
\Te         & 0                 & \Te           & 32612.251248000     \\
\re         & 0.970655035       & \re           & 1.013454000         \\
\Ae         & 37269.126195730   & \Ae           & 53204.256220000     \\
$p_L$       & 4                 & $p_L$         & 3                   \\
$p_R$       & 3                 & $p_R$         & 3                   \\
$N_L$       & 6                 & $N_L$         & 4                   \\
$N_R$       & 8                 & $N_R$         & 8                   \\
$\beta_0$   & 2.292052187       & $\beta_0$     & 2.620396000         \\
$\beta_1$   & -0.019995181      & $\beta_1$     & 0.169768000         \\
$\beta_2$   & 0.198099686       & $\beta_2$     & 0.465289000         \\
$\beta_3$   & 0.231459916       & $\beta_3$     & 0.552208000         \\
$\beta_4$   & -0.136025012      & $\beta_4$     & 0.413921000         \\
$\beta_5$   & -0.639517770      & $\beta_5$     & -5.174652000        \\
$\beta_6$   & -0.277770706      & $\beta_6$     & 9.999998000         \\
$\beta_7$   & 6.274892232       & $\beta_7$     & 3.333865000         \\
$\beta_8$   & -4.898195146      & $\beta_8$     & -9.999764000        \\
\end{tabular}
\end{ruledtabular}
Units: \\
$[T_{\rm e}]     = \textrm{cm}^{-1}$\\
$[A_{\rm e}]     = \textrm{cm}^{-1}$\\
$[r_{\rm e}]     =$ \AA \\
$[\beta_i] =$ \AA$^{-1}$ \\
\end{table}
\subsubsection{Morphing}
\label{subsubsec:morphing}
As discussed above, morphing is another available technique for model refinement. In this case, one does not vary the curve in question directly, but instead varies a morphing function, $f_{\rm m}(r)$, which in turn, scales the \ai curve, $f_{\rm ai}(r)$ \cite{99MeHuxx.methods,99SkPeBo.methods}:
\begin{eqnarray}\label{eq: morphing}
    f_{\textrm{morphed}}(r) = f_{\rm ai}(r)f_{\rm m}(r).
\end{eqnarray}
As was done in \citet{21MiTaTe.NaO}, the \textit{Polynomial decay} morphing function was used: 
\begin{equation}\label{e:f_m}
    f_{\rm m}(r) = \sum^N_{k=0}{B_kz^k(1-\xi_p)+\xi_pB_{\infty}},
\end{equation}

where
$$
    z(r) =  (r-r_{\rm ref})e^{-\beta (r-r_{\rm ref})^2 - \gamma ( r-r_{\rm ref})^4},
$$
$\xi_p$ is as in Eq.~(\ref{eq:surkus}), $B_k$ are variable expansion coefficients, $\beta$ and $\gamma$ are static coefficients typically set to $0.8$ and $0.02$ respectively, $B_\infty$ is typically set to unity to preserve the asymptotic behaviour of $f_{\rm ai}$, $N$ is the order of expansion, and $r_{\rm ref}$ is the expansion center. $r_{\rm ref}$ is set to the equilibrium bond length of the lower energy electronic state. When fitting Eq. \ref{e:f_m}, only the $B_k$ parameters are floated. 

\subsubsection{Constraining Curves}\label{subsec:constrain}
The asymptotic energy limit for all the curves in Fig. \ref{fig:PECs} was constrained by setting the value of $D_{\rm e}^{\rm X}$ to the experimental value of  $D_0$ from \citet{01Joens.OH} with the added offset from the zero-point energy of the \X\ state, $E_0^{\rm X}$ calculated in \Duo, hence
\begin{eqnarray}
    D_{\rm e}^{\rm X} = D_0+E_0^{\rm X}.
\end{eqnarray}
The remaining curves' $D_{\rm e}$ where constrained with respect to $D_{\rm e}^{\rm X}$ such that their respective atomic limit spacings match that of \citet{NISTWebsite}.
\subsection{MARVEL Data Set and Quantum Number Conventions}
\label{subsec:MARVEL}
\citet{jt868} collected 15938 rovibronic transitions from 45 sources and produced values for 1624 empirical rovibronic energy levels for the system of electronic states, \X, \A, \B, \C. 12413 and 1403 of these transitions and empirical energy levels respectively concern the \X\ and \A\  states. 
Table \ref{tab:marvel_coverage} presents the coverage of the MARVEL energy levels used in this study. 
    
The complete set of quantum numbers used to characterise the MARVEL energy levels are the state label (\X, \A,  1~$^2\Sigma^-$, \ldots) total angular momentum, $J$, the vibrational quantum number, $v$, the rotationaless parity, $e/f$ and the projection of the total angular momentum on the molecular axis, $\Omega = \Sigma + \Lambda$, where, $\Sigma$ and $\Lambda$ are projections of the spin and orbital angular momenta on the molecular axis, respectively. 

In line with the Hund's case (a) conventions used in \Duo, the following quantum numbers are used to  characterize the rovibronic states, $|J,~v,~\tau,~\Sigma,~\Lambda,~\Omega,~State\rangle$ where $State$ is a counting number associated with the electronic states as ordered by potential minima, $T_{\rm e}$ and $\tau$ is the state parity $+/-$ (see \citet{jt632} for conversion between $e/f$ and $\tau$), which can be directly related to the MARVEL data set.  

Instead of using the the Hund's case (a) convention, many data sets on the rovibronic state of OH in the literature opt for the rotational quantum number $N$ and the fine structure components, $F_1$ and $F_2$ (Hund's case (b)). In order to correlate these data with our MARVEL data set, their representations have been converted to the rigorous quantum labels $J$, $e/f$  using the following relations.
\\
For the \A\ state: 
\begin{eqnarray}\label{eq:NtoJ}
    J = 
    \begin{cases}
        N+\frac12 & \textrm{for} ~~ F_1\\
        N-\frac12 & \textrm{for} ~~ F_2
    \end{cases}
\end{eqnarray}
\begin{eqnarray}\label{eq:Ftoef}
    e/f = 
    \begin{cases}
        f & \textrm{for} ~~ F_1\\
        e & \textrm{for} ~~ F_2
    \end{cases}
\end{eqnarray}
These relations arise, in particular, as the \A\  state levels are generally presented using Hund's case (b), hence the relation between $J$ and $N$, Eq.~(\ref{eq:NtoJ}) and due to the common approximation,
\begin{eqnarray}
    E_{F_1} =& BN(N+1)+\frac12\gamma N\\
    E_{F_2} =& BN(N+1)-\frac12\gamma (N+1).
\end{eqnarray}
Therefore, because $\gamma \ll B$, \cite{16BrBeWe.OH} where $B$ is the rotational constant and $\gamma$  is the spin-rotation constant,
\begin{eqnarray}
    E_{F_1} > E_{F_2} ~~  \forall ~~ N, J
\end{eqnarray}   
and since the \A\  state has only parity splitting as $\Lambda^{\textrm{A}} = 0$, the $F_1$ and $F_2$ components correspond directly to $f$ and $e$ parities respectively (see \citet{89Herzberg.adhoc_book}). 
\\
For the \X\ state the relevant Hund's case (a) approximation is:
\begin{eqnarray}
    E_{F_1} =& B[(J+S)^2-\Lambda^2] - \frac12(A-2B)\\
    E_{F_2} =& B[(J+S)^2-\Lambda^2] + \frac12(A-2B)
\end{eqnarray}
Where $A$ is the diagonal spin-orbit coupling. This is applied for a given parity and correlates with the following relation \cite{89Herzberg.adhoc_book}:
\begin{eqnarray}
    E_{F_1} =& E\left(\Omega = \frac32\right)\\
    E_{F_2} =& E\left(\Omega = \frac12\right)
\end{eqnarray}
The quantum number correlations for the quartet $1~(^4\Sigma^-,^4\Pi)$ states have not been considered here as there are no experimental data which will require transformation.


\section{Methods: Predissociation Lifetimes}
\label{sec:predissociation}

OH predissociation lifetimes have long been the subject of theoretical and experimental studies, a summary of the available literature data has been provided in Table \ref{tab:prediss_summary}.  
\begin{table*}
\caption{\label{tab:prediss_summary}Summary of predissociation lifetime data available from experimental and theoretical sources.}
\begin{ruledtabular}
\begin{tabular}{lrrrrrrrl}
\multicolumn{1}{l}{}                                & \multicolumn{2}{c}{Lifetime (ps)}                 & \multicolumn{2}{c}{$v$}                             & \multicolumn{2}{c}{$J$}                             & \multicolumn{1}{l}{} &                      \\ \hline
\multicolumn{1}{l}{Reference\footnotemark[1]}       & \multicolumn{1}{l}{Min} & \multicolumn{1}{l}{Max} & \multicolumn{1}{l}{Min} & \multicolumn{1}{l}{Max} & \multicolumn{1}{l}{Min} & \multicolumn{1}{l}{Max} & \multicolumn{1}{l}{Number of Lifetimes} &   Data Type\\ \hline
\multicolumn{8}{c}{Experiment} &                                                                                                                                                                                                                            \\ \hline
05DePoDe \cite{05DePoDe.OH}                         & 17                      & 23                      & 4                       & 4                       & 0.5                     & 7.5                     & 8                                          &Predissociation Lifetimes\\
21SuZhZh \cite{21SuZhZh.OH}                         & 14                      & 130000                  & 2                       & 4                       & 0.5                     & 2.5                     & 2                                          &Predissociation Lifetimes\\
78BrErLy \cite{78BrErLy.OH}                         & 33000                   & 1110000                 & 0                       & 2                       & 0.5                     & 29.5                    & 118                                        &Total Lifetimes\\
91GrFa \cite{91GrFa.OH}                             & 73                      & 500                     & 3                       & 3                       & 0.5                     & 9.5                     & 17                                         &Predissociation Lifetimes\\
92HeCrJe \cite{92HeCrJe.OH}                         & 62.89                   & 312.50                  & 3                       & 3                       & 0.5                     & 13.5                    & 27                                         &Predissociation Rates\\
97SpMeMe \cite{97SpMeMe.OH}                         & 29                      & 167                     & 3                       & 3                       & 4.5                     & 14.5                    & 20                                         &Predissociation Rates\\ \hline
\multicolumn{8}{c}{Theory} &                                                                                                                                                                                                                                  \\ \hline
80SiBaLe \cite{80SiBaLe.OH}                         & 2.95                    & 428.16                  & 3                       & 9                       &                         \multicolumn{2}{c}{Not $J$ Resolved}& 7                                          &Predissociation Lifetimes\\
94KaSa \cite{94KaSa.OH}                             & 3.32                    & 6895.10                 & 1                       & 7                       &                         \multicolumn{2}{c}{Not $J$ Resolved}& 7                                          &Predissociation Lifetimes\\
99PaYa \cite{99PaYa.OH}                             & 6000                    & 9666000                 & 0                       & 4                       & 0.5                     & 30.5                    & 107                                        &Predissociation Lifetimes\\
Grand Total                                         &                         &                         &                         &                         &                         &                         & 313                                        &\\ \hline
\textbf{This work}                                           & 2.1                       & 3600000              & 0                       & 9                       & 0.5                     & 35.5                    & 374                                        &Predissociation Lifetimes\\
\end{tabular}
\end{ruledtabular}
\footnotetext[1]{Reference tags are given as YYAaBbCc, where YY is the last two digits of the publication year, AaBbCc are first two letters of (up to) first three authors surnames in order of appearance.}
\end{table*}
The data provided by \citet{92HeCrJe.OH, 97SpMeMe.OH} are given as predissociation rates and these have been inverted to return the required predissociation lifetimes. \citet{78BrErLy.OH} provides measurements of the total lifetime which is given by 
\begin{eqnarray} \label{eq:tau_tot}
    \tau = \frac{\tau_r \tau_p}{\tau_r+\tau_p}
\end{eqnarray}
where $\tau_{r}$ and  $\tau_{p}$ are the radiative and predissociative lifetimes respectively. As these values include also the radiative lifetime, we expect our calculations to be greater than these values. This is compounded by the fact that the predissociation in the region of $v\in[0,1,2]$ is less dominant than in $v\geq3$. 

A novel method for the calculation of predissociation lifetimes is applied in this section. The full details of the method will be available in Paper I, however, a brief summary of the method is given below. 

\subsubsection{Energy resonances}
\label{subsec:stabilization}

An excited rovibronic state from \A\  can predissociate via a spin-orbit interaction with a local repulsive state inducing the decay  \cite{21SuZhZh.OH,87LeFr.OH,95Le2.OH,99PaYa.OH,11LiZh.OH}.  

\begin{figure}[H]
\includegraphics[width=0.5\textwidth]{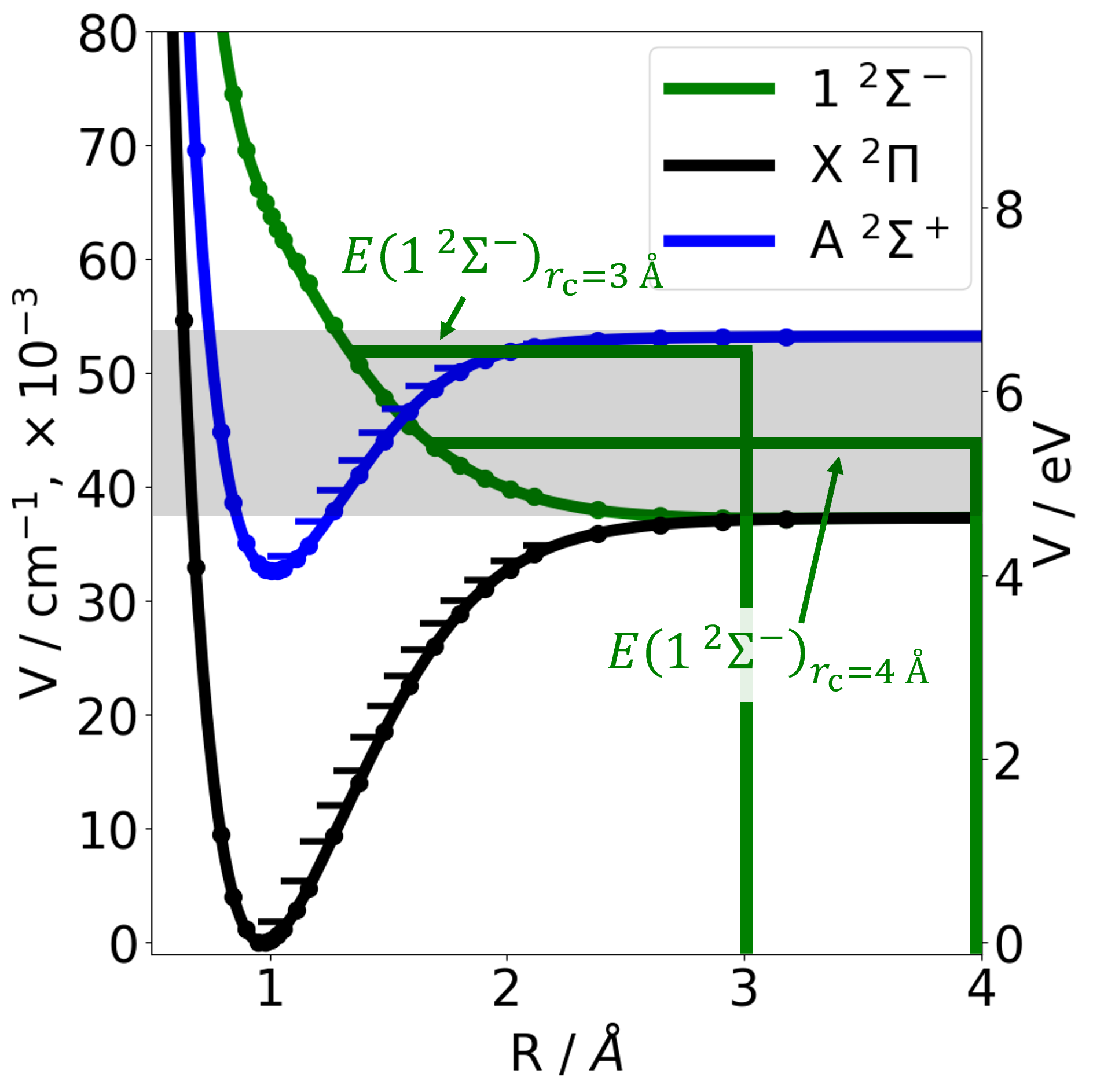}
\caption{\label{fig:toy_model} A toy model of OH consisting of the \X, \A, and \IS\ electronic states. The two green levels, $E(1~^2\Sigma^-)_{r_{\rm c}=3,4~\rm \mathring{A}}$  are continuum levels in the \IS\ electronic state with the same quantum numbers, however with different right-side box limits, \rc. This highlights the box size dependence of repulsive state energy levels. The shaded region covers energy levels between $D_{\rm e}^{\rm X}$ and $D_{\rm e}^{\rm A}$.}
\end{figure}

In Fig. \ref{fig:toy_model}, the two green horizontal lines illustrate unbound state energy levels retrieved from \Duo\ calculations with different right-side box limits, \rc. Both sets of calculations are performed with the same left-side box limit $r_{\rm min}$. By construction the nuclear motion wavefunctions are only finite between \rc\ and $r_{\rm min}$. 
The two green unbound state energy levels have the same quantum numbers $J, v, \tau, \Sigma, \Lambda, \Omega$, however, $E_{r_{\rm c} = 3 \rm \mathring{A}}$ returns a higher term value than $E_{r_{\rm c} = 4 \rm \mathring{A}}$.  This is akin to the particle in a box, where increasing the size of the box compresses the distribution of energy levels closer to zero. Repeated calculations over varying \rc\ in \Duo\ for iso-numeric continuum energy levels exhibits a relationship where the term values decrease with \rc, this is referred to as the \textit{energy profile} of the continuum energy level.

The spin-orbit interaction between the \IS and \A\  states, $\langle ^2\Sigma^-|\xi|^2\Sigma^+\rangle$, hence displaces the \A\  state levels, which would otherwise be stable under variation of \rc. This \textit{induced energy profile}, allows us to study the predissociation characteristics of the quasi-bound \A\  state levels. 

The energy levels of interest for this study are found in the shaded region of Fig. \ref{fig:toy_model}. This covers 424 rovibronic energy levels in the \A\  state.

The quantum number coverage for the relevant energy levels is presented in Table ~\ref{tab:prediss_coverage}. Being above the first dissociation channel ($D_{\rm e}^{\rm X}$), each of these levels is quasi-bound and meta-stable with an associated characteristic lifetime, which we compute here. As discussed below, not all levels in the range had calculable predissociation lifetimes as per the method in Paper I, especially for levels with energy close to $D_{\rm e}^{\textrm{X}}$. 

\begin{table}
\begin{ruledtabular}
\caption{\label{tab:prediss_coverage}Quantum number coverage for quasi-bound \A\ state rovibronic levels. }
\begin{tabular}{rrrrr}
\multicolumn{1}{c}{v} & \multicolumn{1}{c}{$J_{\rm min}$} & \multicolumn{1}{c}{$J_{\rm max}$} & \multicolumn{1}{c}{Number of Levels} & \multicolumn{1}{c}{Number of Lifetimes \footnotemark[1]{}} \\ \hline
0                     & 13.5                     & 35.5                     & 44                                  & 21                                \\
1                     & 2.5                      & 33.5                     & 62                                   & 35                                \\
2                     & 0.5                      & 31.5                     & 63                                   & 63                                \\
3                     & 0.5                      & 28.5                     & 57                                   & 57                                \\
4                     & 0.5                      & 25.5                     & 51                                   & 51                                \\
5                     & 0.5                      & 22.5                     & 45                                   & 45                                \\
6                     & 0.5                      & 19.5                     & 39                                   & 39                                \\
7                     & 0.5                      & 15.5                     & 31                                   & 31                                \\
8                     & 0.5                      & 11.5                     & 23                                   & 23                                \\
9                     & 0.5                      & 4.5                       & 9                                     &  9                                 \\
Total              &                            &                              & 424                                & 374                              
\end{tabular}
\end{ruledtabular}
\footnotetext[1]{Number of levels for which a lifetime calculation is meaningful.}
\end{table}

Since repulsive electronic states have an infinitely dense  set of continuum states, the position of the predissociating state, \AState, where $Q$ is the complete set of quantum numbers, \quantumnumbers, 
has an uncertainty in the energy caused by the Pauli exclusion principle. This infinite set of states cannot be recovered using bound state methods for solving the Schr\"{o}dinger equation for $r_{\rm c} \rightarrow \infty$, such as used by \Duo, although more expensive scattering methods can be employed using \Duo\ just for the inner region\cite{jt755}. 

Here we  approximate the continuum behaviour by discretizing the continuum \cite{jt840} and adapting the stabilization method \cite{70HaYaHo.adhoc,93MaRaTa.adhoc,82BaSi.adhoc} to characterize the quasi-bound states. In this method, infinite potential walls are assumed at the box limits $r_{\rm min}$ and $r_{\rm c}$, leading to bound-like rovibronic solutions even for continuum or quasi-bound states. By changing the position of the right-side wall $r_{\rm c}$, different continuum state term values can be generated and the uncertainty of the position of the quasi-bound states can be quantified as follows.

For a given box size defined by $r_{\rm c}-r_{\rm min}$  with a infinite potential wall,  one can compute the term values  $\tilde{E}_{|\textrm{A},Q\rangle}|_{r_{\rm c}}$ of the given set of quasi-bound, meta-stable states $\{$\AState$\}$. By varying $r_{\rm c}$,  one can then establish a dependence of $E_{|\textrm{A},Q\rangle}$ on $r_{\rm c}$. In this study, $r_{\rm c}$ was varied between $r_{\rm c} = 8$ and $8.999$ \AA~with the uniform spacing of 999 values ($\delta r_{\rm c} = 0.001$~\AA). This was done by repeating the \Duo~calculations while varying the right-side limit, \rc (see Sec. \ref{subsec:global}) and extracting the eigenenergies at each step. The number of the \Duo\ sinc-DVR \cite{92CoMixx.method} grid points was increased as the box size was increased to maintain a uniform grid. 
Out of 999 
runs, about 20 calculations failed, but these were simply ignored.

\begin{figure}[H]
\includegraphics[width=0.5\textwidth]{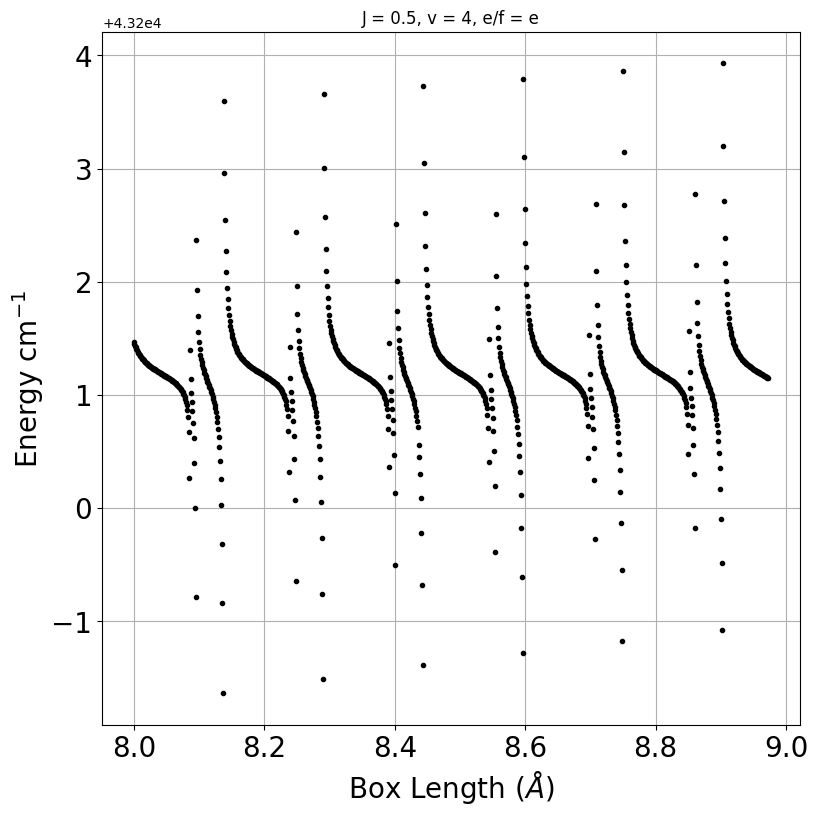}
\caption{\label{fig:evl1} The term value of the $|\textrm{A} ^2\Sigma^+, J=0.5, v = 4, e \rangle$ state as a function of the box length of the calculation in \Duo, the variation of the energy is induced by crossings with energy levels in the repulsive states \IS, \IQS, and \IQPi. }
\end{figure}

\begin{figure}[H]
\includegraphics[width=0.5\textwidth]{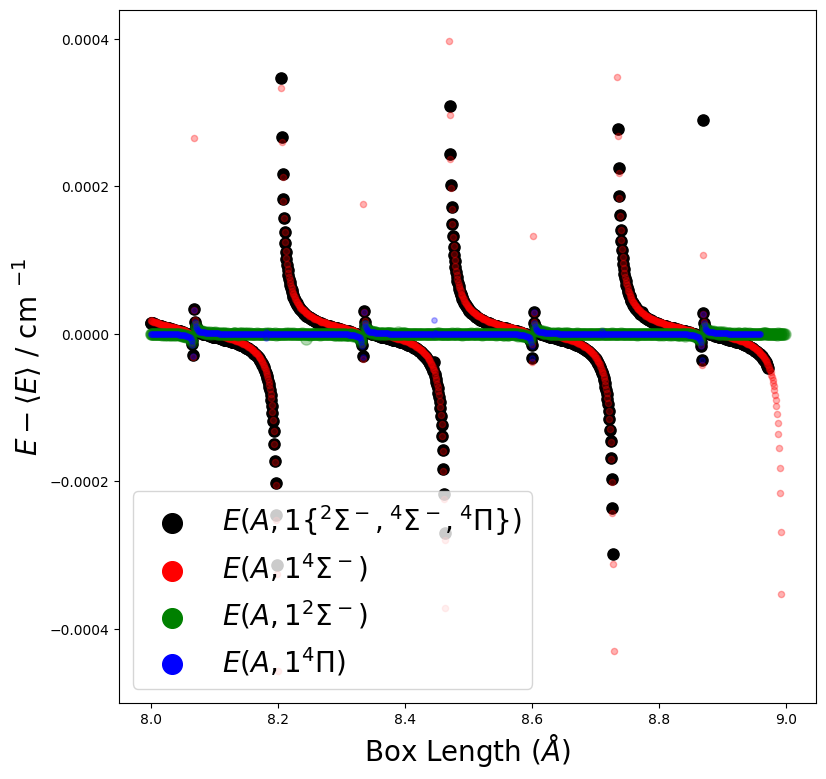}
\caption{\label{fig:layering}Term value of the $|\textrm{A} ^2\Sigma^+, J=0.5, v = 2,  e \rangle$ state as a function of the box length of the calculation in \Duo. The different colors correspond to different spin-orbit contributions. 
}
\end{figure}

Figures~\ref{fig:evl1} and \ref{fig:layering} illustrate  the box-size dependence for two \A\  quasi-bound states of OH, $|\textrm{A} ^2\Sigma^+, J=0.5, v = 4, e \rangle$ and $|\textrm{A} ^2\Sigma^+, J=0.5, v = 2, e \rangle$ showing their  `stabilisation' character. The behavior is seen to be periodic with discontinuities at visually regular intervals and a central energy region in the middle. 
The typical resonance-like shapes are due to the interactions with continuum states.  The discontinuities occur at the geometries where a crossing continuum energy level goes from pushing the quasi-bound level down to pushing it up and provide the energy level resonances from which we can calculate lifetimes. The line broadening parameter, $\Gamma$ and hence the lifetime of the state, $\tau \propto 1/\Gamma$, are associated with the widths of these resonances.

The interaction between \A\ and the three continuum states 1~$^2\Sigma^-$, 1~$^4\Sigma^-$, 1~$^4\Pi$  are through the corresponding spin-orbit couplings from Fig.~\ref{fig:SOCs}. There are several structures visible in Fig. \ref{fig:evl1}, the asymptotes with the large wings and the asymptotes with the small wings. These substructures are caused by spin-orbit interactions with different electronic states, $1\,^2\Sigma^-$, $1\, ^4\Sigma^-$, or $1\,^4\Pi$. The individual contributions can be resolved by performing  calculations with only one of the three repulsive states and associated spin-orbit couplings present at a time, as illustrated in Fig.~\ref{fig:layering} for the $|\textrm{A} ^2\Sigma^+, J=0.5, v = 2,  e \rangle$ state (this state has been chosen for its simplicity). It shows that all three states contribute to the minor asymptotic wings but only the \IQS \ state corresponds to the major wings. This makes the \IQS \ state a bigger contributor to the predissociative decay for this particular state (due to a greater effect on the overall broadening), which is corroborated by the  branching ratios reported by \citet{11LiZh.OH}  who show that  the \IQS\ state is the primary branch. 

\begin{figure}[H]
\includegraphics[width=0.5\textwidth]{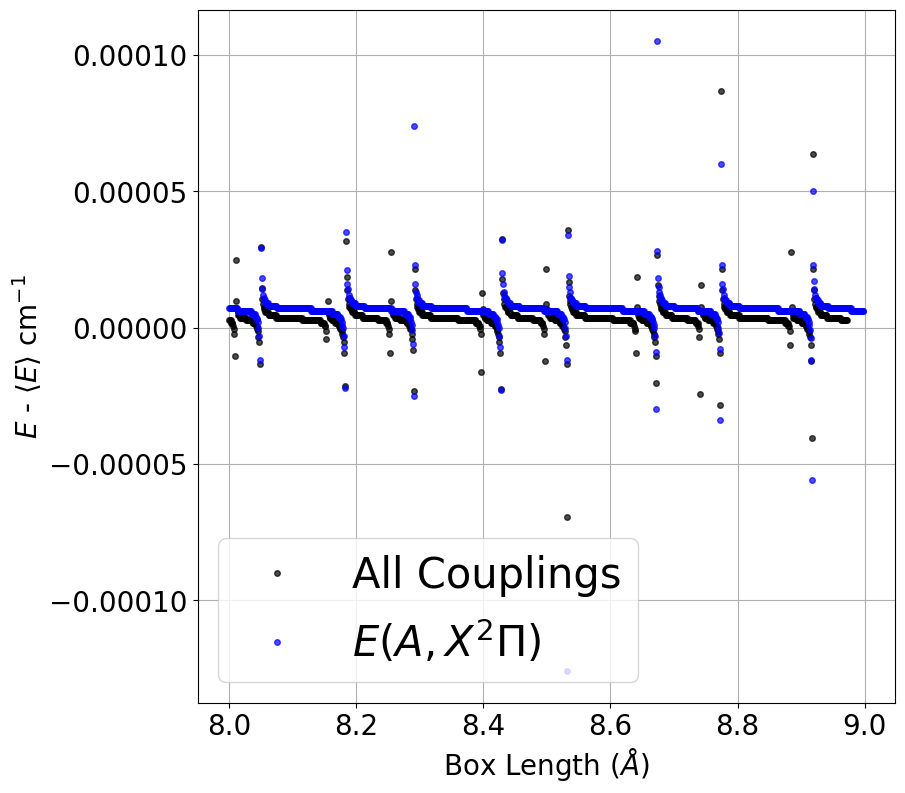}
\caption{\label{fig:OnlyX} Term value of the $|\textrm{A} ^2\Sigma^+, J=13.5, v = 1, f \rangle$ state as a function of the box length for a \Duo\ calculation including the \X\ and \A\  states only. The blue markers represent a shift in energy induced by the spin-orbit coupling between the \X\  and \A\  states on the \A\  state level.}
\end{figure}

For energy levels above $D_{\rm e}^{\rm X}$, \X\ state rovibronic levels become unbound and contribute to the perturbation of the \A\ state level energies and hence provide an additional avenue for predissociation through the ground electronic state, see Fig. \ref{fig:OnlyX}. The effect of predissociation through the \X\  state is small, and does not appear to have previously been reported. This effect cannot be isolated in our calculations, however, as removing the spin-orbit coupling between the \X\  and the \A\  states shifts the energy levels too strongly (hundreds of \cm) to be comparable. 

\subsubsection{Statistical treatment}
\label{sec:stat_treat}
\begin{figure}[H]
\includegraphics[width=0.5\textwidth]{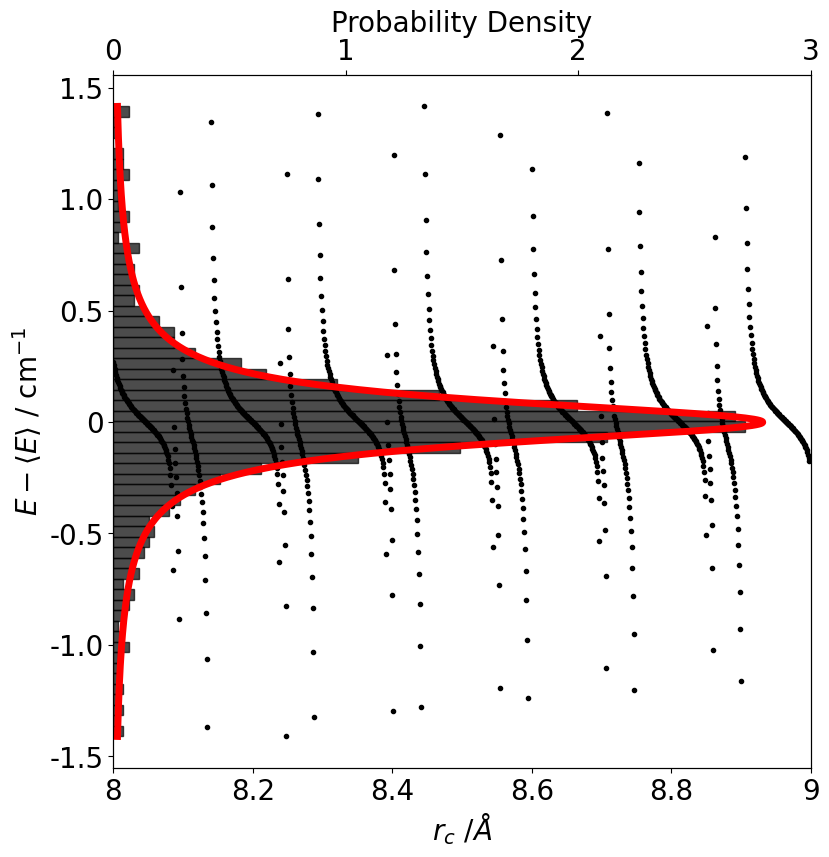}
\caption{\label{fig:total} Plot of the energy against box length with the fitted histogram of energy points for the $|\textrm{A} ^2\Sigma^+, J=0.5, v = 4, e \rangle$ state.}
\end{figure}

To calculate the predissociation lifetime from the above results, one calculates the integral normalised histogram of the energy data points centered around the mean $(E-\langle E\rangle)$, see Fig. \ref{fig:total} for an example. The binning of the histogram and the training of the Lorentzian parameters is non-trivial and is discussed in Paper I.

The probability function describing the system's energy profile has the Lorentzian distribution \cite{95Le1.OH}:
\begin{eqnarray}
    \label{eq:loren}
    L(\tilde{E};\tilde{E}_0,\Gamma) = \frac{\frac{1}{\pi}( \frac{1}{2} \Gamma )}{(\tilde{E}-\tilde{E}_0)^2+(\frac12\Gamma)^2}
\end{eqnarray}
where $\tilde{E}_0$  is the peak position of the Lorentzian and $\Gamma$ is the full-width at half-maximum (line broadening parameter, in \cm). The parameters $\{\tilde{E}_0, \Gamma\}$ can be trained to best represent the histogram in Fig. \ref{fig:total}, then the parameter $\Gamma$ is inverted via the relationship 

\begin{eqnarray}
    \tau_{\rm p} = \frac{1}{2\pi c \Gamma}
\end{eqnarray}
where $c$ is the speed of light in $\textrm{cm s}^{-1}$. 

There is an associated uncertainty to the fitting of the Lorentzian profile and this is discussed in Sec. \ref{subsubsec:Uncertainty Estimation}. The \Python\ package, \binslt~was written to perform these calculations; it is available for download from  \href{https://github.com/exomol}{https://github.com/exomol}. 

A visualisation of the Lorentzian-distributed energy histogram is shown in Fig. \ref{fig:total}, where it is superimposed with the corresponding induced energy profile for the $|\textrm{A} ^2\Sigma^+, J=0.5, v = 4, e \rangle$ state .

\subsubsection{Uncertainty Estimation}\label{subsubsec:Uncertainty Estimation}

The largest uncertainties arise in states with long predissociation lifetimes. Here we will consider only the lifetimes which are shorter than the radiative lifetimes. As discussed in Paper I, the uncertainty is made up for four components, which are assumed independent and systematic and so are added to produce a final uncertainty $\Delta_{\tau}$
\begin{eqnarray}
    \Delta_{\tau} = \Delta_M + \Delta_R + \Delta_B + \Delta_C, 
\end{eqnarray}
where $\Delta_M$ is the uncertainty from convergence which is defined as $1-M$ where $M$ is the convergence level of the lifetime for a given state. This is typically the largest uncertainty and in cases of long lifetimes ($\tau_p > 10^5$ ps) this can be as high as 20\%. For lifetimes between $10^4$ and $10^5$ ps, $\Delta_M$ is evaluated at 10\% and for lifetimes shorter than $10^4$ ps, this is found to be 5\%. 

$\Delta_R$ and $\Delta_B$ is the uncertainty from the positions of the repulsive and bound states' energy levels. $\Delta_B$ is very difficult to measure directly, as it requires that the energy level positions of the bound states can be controlled with high precision in order to establish a relationship between the energy offset and the lifetime. This is, at the moment, infeasible. This has been estimated, however, by probing the effect the repulsive energy level positions have on the lifetimes. The asymptotic energy of the continuum states go to the O($^3$P) level, the same as the \X\ state. $D_{\rm e}(\textrm{X})$ has been experimentally measured to an uncertainty of 10 \cm\  by \citet{01Joens.OH}  (see sec. \ref{subsec:constrain}). Our continuum curves were shifted down by 20 \cm\   and the lifetimes for the predissociative states with $J \in [0.5, 1.5]$ were re-computed and compared to the original model. This gives a conservative estimate on the uncertainty due to the position of the continuum curves, $\Delta_R$ of about 5\%. The \A\  state energy level position induced uncertainty, $\Delta_B$ is hence also estimated at 5\%

\begin{figure}[H]
\includegraphics[width=0.5\textwidth]{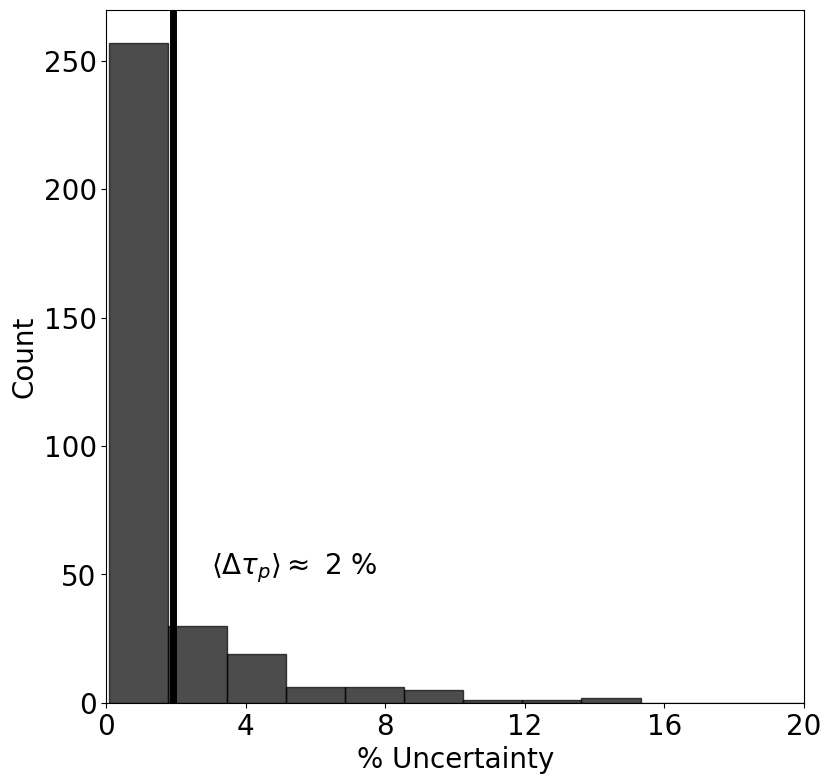}
\caption{\label{fig:fitting_unc} Distribution of percentage uncertainties in the lifetime induced by the computational procedure.}
\end{figure}

$\Delta_C$ is the uncertainty of the numerical procedure excluding the problem of convergence. This is evaluated \textit{in situ} for each state and has an average value over all states of approximately 2\% in the case of OH. These uncertainties are distributed exponentially as is seen in Fig. \ref{fig:fitting_unc}. 

The lifetimes supplied in the supplementary material have been fully treated as in the methods described above and more completely, in Paper I. Tab. \ref{tab:uncertainty_rules} gives a summary of the average uncertainties over different ranges of lifetimes in ps.

\begin{table}[h]
\centering
\caption{\label{tab:uncertainty_rules} Rules for attributing uncertainties for predissociation lifetimes calculated using the stabilization method for OH}
\begin{tabular}{lrrr}
Quantity            & \multicolumn{3}{l}{Value}            \\ \hline
Lifetime range / ps & $\left[0,10^{4}\right]$ & $\left[10^{4},10^{5}\right]$ & $\left[10^{5},10^{7}\right]$   \\ 
Average Uncertainty / \%& 15+$\Delta_C$& 20+$\Delta_C$& 30+$\Delta_C$\end{tabular}
\end{table}


\section{Results}
\label{sec:results}
\subsection{Spectroscopic model parameters}
Refinement of the curves  representing the coupled model of the two states \X\  and \A\  in \Duo~was performed by floating the parameters $r_{\rm e}({\rm X},{\rm A})$, $T_{\rm e}({\rm A})$, $\beta_i({\rm X,A})$ (PECs) and $B_k$ (spin-orbit and  L-uncoupling curves)

Couplings to higher electronic states have not been considered for sake of simplicity in the model. Furthermore, couplings to repulsive electronic states for refinement of bound state energy levels have not been considered either. This has been done due to computational limitations as the box size of the calculation becomes an independent variable for the energy levels, see discussions above. 

Along with the adjusted PECs and the \textit{morphed} coupling curves, three more \textit{empirical}, diagonal curves have been introduced to account for the aforementioned missing couplings \cite{jt632}. These are the spin-rotation, $\gamma(X,A)$, the $\Lambda$-doubling $\alpha_{p2q}(X)$ and $q(X)$ \cite{87BrChMe.adhoc}, and the Born-Oppenheimer breakdown (BOB) radial property function \cite{02LeHu.adhoc}. The inclusion of such factors was necessary to refine the energy level reproduction without requiring the introduction of further electronic states or off-diagonal couplings. The \v{S}urkus polynomial expansion function, $C(r)$ (see Eq.~(\ref{eq:bobleroy})), has been used as the parametric representation \cite{LEVEL},
\begin{eqnarray}
    \label{eq:bobleroy}
    C(r) = T_{\rm e} + (1-\xi_p)\sum_{i\geq0}B_i\xi_p^i + \xi_pB_{\infty} 
\end{eqnarray}
where $\xi_p$ is the reduced coordinate in Eq.~(\ref{eq:surkus}). The parameters $B_{i}$ and $B_{\infty}$ 
are floated.

The dissociation limits, $D_{\rm e}(\textrm{X,A})$ were fixed to the values 
$D_{\rm e} = D_0 + \textrm{ZPE}$, where 
the zero point energy of the ground state was extracted from the \Duo~ calculation and found to be $\sim$ 1918 \cm, $D_0(\textrm{X})$ was measured by \citet{01Joens.OH} to $\pm10$ \cm\ $D_{\rm e} = D_0 + \textrm{ZPE}$, while $D_{\rm e}(\textrm{A})$ was estimated by adding the difference in energy between the O($^3$P) and O($^1$D) atomic levels using values from the NIST Online Database \cite{NISTWebsite}.

In \Duo, one can fit spectroscopic model both  against data sets of energy levels and of transitions in \Duo. The process taken in this work was to fit it against the experimentally derived (MARVEL) energy levels by \citet{jt868}. 

A summary of fitted parameters is available in Tables \ref{tab:fitted_parameters_PEC}, \ref{tab:fitted_parameters_morphing}, \ref{tab:fitted_parameters_empirical} and a summary of uncertainties and RMS values is available in table \ref{tab:RMS}. The \Duo\ input file is given in the supplementary information. 

\begin{table}[H]
\caption{\label{tab:fitted_parameters_PEC} Summary of the PEC parameters obtained from least squares fitting the \X\  and \A\  states of OH to MARVEL energy levels from \citet{jt868} in Duo \cite{Duo}. All the figures are kept for the sake of preserving  accuracy and reproducibility.}
\begin{ruledtabular}
\begin{tabular}{lrr}
\multicolumn{1}{c}{Parameter} & \multicolumn{1}{c}{\X} & \multicolumn{1}{c}{\A} \\ \hline
\Te                           & 0& 32663.976309068300
\\
\re                           & 0.970020962666
& 1.005285998312
\\
\Ae                           & 37501.792600000000
& 53369.654600000000
\\
$p_L$                         & 4& 3\\
$p_L$                         & 3& 3\\
$N_L$                         & 6& 4\\
$N_R$                         & 9& 8\\
$\beta_0$                     & 2.287411349533
& 2.625193503966
\\
$\beta_1$                     & -0.008481205971
& 0.137874402944
\\
$\beta_2$                     & 0.138199166549
& 0.280685923916
\\
$\beta_3$                     & -0.066491288959
& 0.687146152607
\\
$\beta_4$                     & 0.249817374269
& 1.374869049258
\\
$\beta_5$                     & 1.855331479551
& -2.136036385410
\\
$\beta_6$                     & 2.083877071440
& -14.858091727498
\\
$\beta_7$                     & -21.177505781638
& 45.312091462135
\\
$\beta_8$                     & 34.392719961367
& -32.477114187491
\\
$\beta_9$                     & -16.859334102231
&                      
\end{tabular}
\end{ruledtabular}
\end{table}

\begin{table}[H]
\caption{\label{tab:fitted_parameters_morphing} Summary of the \textit{morphing} parameters obtained from least squares fitting the \X\  and \A\  states of OH to MARVEL energy levels and transitions from \citet{jt868} in Duo \cite{Duo}.}
\begin{ruledtabular}
\begin{tabular}{lrrr}
\multicolumn{1}{c}{Parameter}  & $\xi_X$           & $\xi_{XA}$           & $L_{XA}$         \\ \hline
\re                            & 0.970655034638
& 0.970655034638&  0.9704443874981\\
$\beta$                        & 0.8
& 0.8          & 0.8         \\
$\gamma$                       & 0.02         & 0.02         & 0.02        \\
$p$                            & 6            & 6            & 2           \\
$B_0$                          & 1.008496273609
& 0.675170320954
& 0.772527250694
\\
$B_1$                          & -0.504517931107
& -0.799178556532
& -0.649959182912
\\
$B_2$                          & 0.127429582028
& -2.036097977961
& -0.712564567820
\\
$B_{\infty}$                   & 1.000000000
& 1.000000000
& 1.000000000
\end{tabular}
\end{ruledtabular}
\end{table}

\begin{table*}
\caption{\label{tab:fitted_parameters_empirical} Summary of the empirical parameters obtained from least squares fitting the \X\  and \A\  states of OH to MARVEL energy levels and transitions from \citet{jt868} in Duo \cite{Duo}}
\begin{ruledtabular}
\begin{tabular}{lrrrrr}
\multicolumn{1}{c}{} & \multicolumn{1}{c}{BOB(\A)} & $\gamma_X$    & $\gamma_A$           & \multicolumn{1}{c}{$\alpha_{p2q}$} & \multicolumn{1}{c}{$q$}   \\ \hline
\re(\AA)            & 0.969785542585& 0.969785542585& 0.969785542585& 0.969785542585& 0.969785542585\\ 
$p$                  & 2& 2& 2& 2& 2\\
$N$                  & 3& 3& 3& 3& 3\\
$B_0$  (\AA$^{-1}$)             & -0.013387427200& -0.082615835520& 0.107579056717& 0.081298271720& -0.015657399328\\
$B_{\infty}$ (\AA$^{-1}$)         & -0.000897768443& 0.029822159900& 0.019259293154& -0.116865046681& 0.016057094058
\end{tabular}
\end{ruledtabular}
\end{table*}

\begin{table}[H]
\caption{\label{tab:RMS} Summary of the energy RMS values, in \cm, obtained from least squares fitting the \X\  and \A\  states of OH to MARVEL transitions from \citet{jt868} in Duo \cite{Duo}}
\begin{ruledtabular}
\begin{tabular}{crr}
v           & \multicolumn{1}{c}{${\rm RMS}_{\rm X}$} & \multicolumn{1}{c}{${\rm RMS}_{\rm A}$} \\ \hline
0           & 0.08                       & 0.07                       \\
1           & 0.09                       & 0.19                       \\
2           & 0.15                       & 1.05                       \\
3           & 0.08                       & 0.76                       \\
4           & 0.12                       & 0.61                       \\
5           & 0.06                       & 1.96                       \\
6           & 0.11                       & 3.23                       \\
7           & 0.14                       & 4.70                       \\
8           & 0.08                       & 1.64                       \\
9           & 0.25                       & 4.58                       \\
10          & 0.25                       &                            \\
11          & 0.70                       &                            \\
12          & 1.34                       &                            \\
13          & 1.13                       &                            \\
Whole State & 0.33                       & 1.79                      
\end{tabular}
\end{ruledtabular}
\end{table}

\subsection{Predissociation lifetimes}

\binslt~was used to compute the predissociation lifetimes of 374 quasi-bound \A\ rovibronic states with follow up processing completed as further described in Paper I. 50 out of the allowed 424 states could not be processed as their energy profiles were too narrow and had insufficient resolution to compute $\Gamma_{\rm p}$.

\begin{figure}[H]
    \centering
    \includegraphics[width=0.5\textwidth]{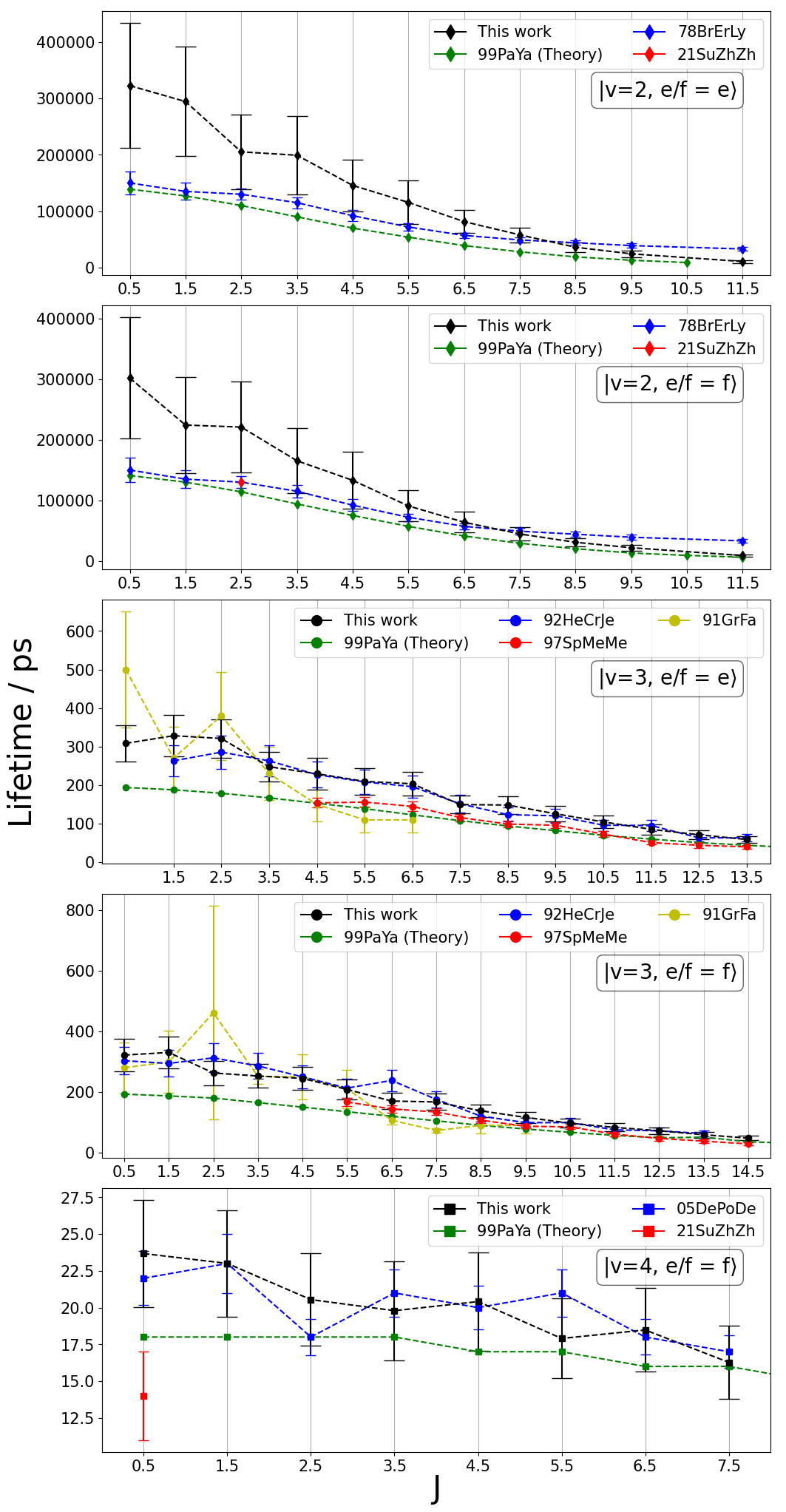}
    \caption{\label{fig:comparison} Predissociation lifetimes as calculated in this work compared to experimental values from 05DePoDe\cite{05DePoDe.OH}, 78BrErly \cite{78BrErLy.OH}, 91GrFa \cite{91GrFa.OH}, 92HeCrJe\cite{92HeCrJe.OH}, 97SpMeMe\cite{97SpMeMe.OH}, and 21SuZhZh\cite{21SuZhZh.OH}, and theoretical values from 99PaYa \cite{99PaYa.OH}.}
\end{figure}

A comparison with the lifetimes available in the literature (experiment and theory) and this work is illustrated in Fig. \ref{fig:comparison}, where the lifetime values are plotted as a function of $J$. The corresponding $J$-averaged percentage errors and their standard deviations are given in Table \ref{tab:OC}. A full tabulation of these data is available in the supplementary material for reference.

Table \ref{tab:OC} shows that  there is a satisfactory agreement between the lifetimes presented here and experiment; there is particularly good agreement for the cases where $v=3,4$, see Fig. \ref{fig:comparison}. In these cases, most results agreed within uncertainty (51/73 cases) 
and in regions where they did not agree, the mean percentage real difference in this region is $15\%$. The percentage real difference, $^{\rm R}\Delta_{\%}$ and percentage difference, $\Delta_{\%}$ here is defined as
\begin{eqnarray}
    D = & \frac{|\Delta|-(\delta_{\rm calc}+\delta_{\rm lit})}{\tau_{\rm lit}}\\
    ^{\rm R}\Delta_{\%} = & 
    \begin{cases}
        D & \textrm{for} ~~ D > 0\\
        0 & \textrm{otherwise}
    \end{cases}\\
    \Delta_{\rm \%} =& \frac{\tau_{\rm lit}-\tau_{\rm calc}}{\tau_{\rm lit}} \equiv \frac{\Delta}{\tau_{\rm lit}},
\end{eqnarray}
where $\tau_{\rm lit, calc}$ are the literature and calculated predissociation lifetimes respectively, $\Delta$ is the difference in those values, and $\delta_{\rm lit, calc}$ are the uncertainties quoted in the literature and calculated values respectively. 
 
The noticeable exceptions to this are the lifetimes for $v\in[0,1,2]$ from \citet{78BrErLy.OH}. The experimental values in question here are total lifetimes rather than predissociation lifetimes. For $v<2$, radiative decay is the dominant source of broadening, hence the lifetimes of \citet{78BrErLy.OH} measurements should  underestimate  the predissociation lifetime. For $v=2$, predissociation is dominant, however radiative decay appears to be significant in the low $J$ region ($<6.5$). 

One can consider the individual widths of the radiative and predissociative components, $\Gamma_{\rm r,p}$ and that the total width is $\Gamma = \Gamma_{\rm r} + \Gamma_{\rm p}$ and extract the dominance of the radiative decay, $R_D$ such that
\begin{eqnarray}
    R_D = \frac{\Gamma_{\rm r}}{\Gamma}
\end{eqnarray}
From our calculations, we compute $R_D$ and estimate the predissociation lifetime of \citet{78BrErLy.OH} through the relation
\begin{eqnarray}
    \tau_{\rm p} = \frac{\tau}{1-R_D}
\end{eqnarray}
and from this we are able to recover agreement within uncertainty with \citet{78BrErLy.OH}. 

\begin{table}[H]
\begin{ruledtabular}
\caption{\label{tab:OC} Summary of comparison between predissociation lifetimes as calculated in this work and available literature values. Differences are averaged over $J$ for a given $v$. Only rotationally resolved literature lifetimes are compared here (see Table \ref{tab:prediss_summary}). Full comparison dataset provided in supplementary data. }
\begin{tabular}{llcrccr}
\multicolumn{1}{l}{Type} & \multicolumn{1}{l}{Ref}      & \multicolumn{1}{l}{v} & \multicolumn{1}{l}{$\langle \Delta_{\%}\rangle$} & \multicolumn{1}{l}{Number} & \multicolumn{1}{l}{$C_{\rm A}$} & \multicolumn{1}{l}{$\langle ^{\rm R}\Delta_{\%}\rangle$} \\ \hline
\multirow{8}{*}{Exp}        & \multicolumn{1}{l}{05DePoDe\cite{05DePoDe.OH}}& 4                     & -0.2                                                 & 8                           & 8                                       & 0.0                            \\ \cline{2-7} 
                            & \multirow{2}{*}{21SuZhZh\cite{21SuZhZh.OH}}   & 2                     & -70.0                                                & 1                           & 0                                       & 12.2                           \\
                            &                                               & 4                     & -69.1                                                & 1                           & 0                                       & 21.7                           \\ \cline{2-7} 
                            & \multirow{3}{*}{78BrErLy\cite{78BrErLy.OH}}   & 0                     & -72.6                                                & 9                           & 5                                       & 77.0                           \\
                            &                                               & 1                     & -137.5                                               & 6                           & 3                                       & 123.1                          \\
                            &                                               & 2                     & -28.5                                                & 22                          & 12                                      & 24.2                           \\
 \cline{2-7}& 91GrFa\cite{91GrFa.OH}& 3& -26.7& 17& 13&39.5\\ \cline{2-7} 
                            & \multicolumn{1}{l}{92HeCrJe\cite{92HeCrJe.OH}}& 3                     & -2.0                                                 & 27                          & 26                                      & 2.1                            \\ \cline{2-7} 
                            & \multicolumn{1}{l}{97SpMeMe\cite{97SpMeMe.OH}}& 3                     & -40.5                                                & 20                          & 4                                       & 11.7                           \\ \hline
\multirow{5}{*}{Theory}     & \multirow{5}{*}{99PaYa\cite{99PaYa.OH}}       & 0                     & -76.7                                                & 11                          & 4                                       & 34.9                           \\
                            &                                               & 1                     & -77.1                                                & 9                           & 3                                       & 27.0                           \\
                            &                                               & 2                     & -87.2                                                & 23                          & 0                                       & 33.0                           \\
                            &                                               & 3                     & -48.0                                                & 35                          & 1                                       & 25.6                           \\
                            &                                               & 4                     & -15.3                                                & 21                          & 12                                      & 6.1                            \\ 
\end{tabular}
\end{ruledtabular}
The columns are as follows:
Type = Experiment or Theory\\
Ref = Reference\\
$v$ = Vibrational quantum number\\
$\langle\Delta_{\%}\rangle$ = Average percentage difference \\
Number = number of $J$ resolved levels\\
$C_A$= Number of levels whose lifetimes agree within uncertainty\\
$\langle ^{\rm R}\Delta_{\%}\rangle$: Average $^{\rm R}\Delta_{\%}$  for values which do not agree within uncertainty
\end{table}

\begin{figure}[H]
\includegraphics[width=0.5\textwidth]{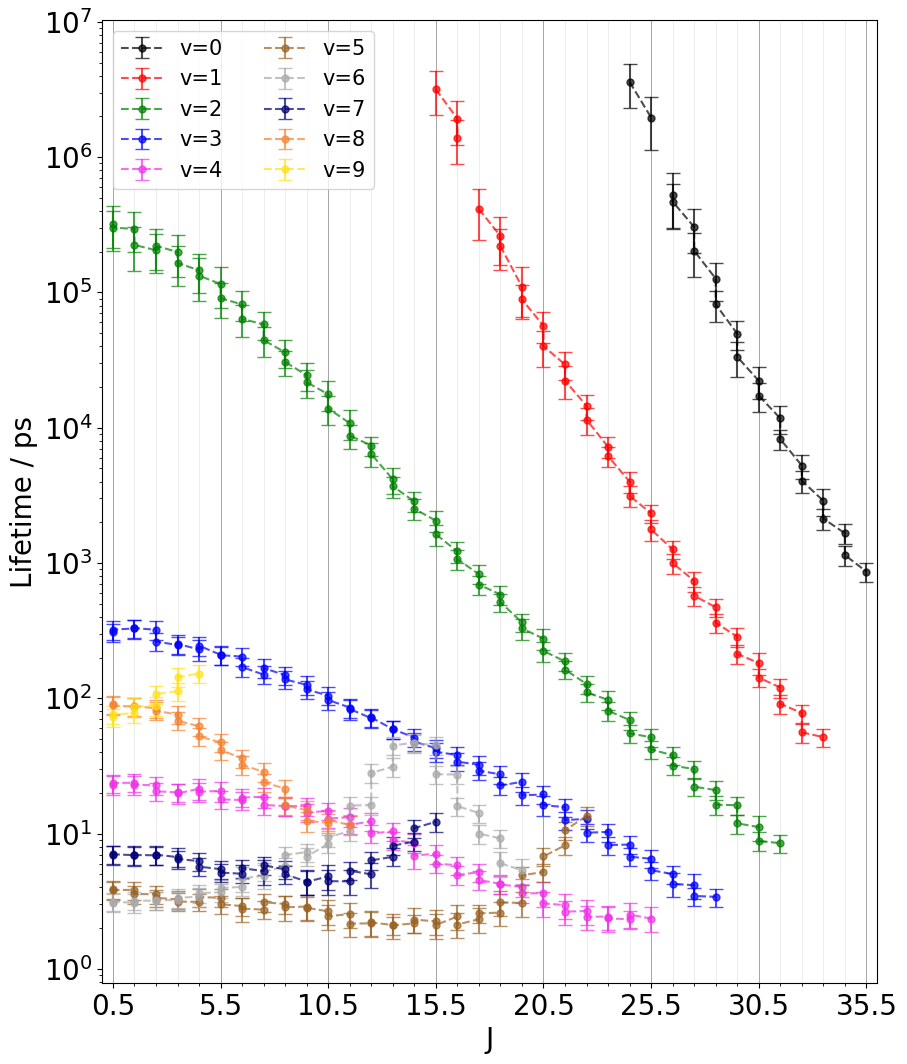}
\caption{\label{fig:ltJ} Statistical predissociation lifetimes calculated for quasi-bound rovibronic \A\  state levels using code, \binslt~from Paper I. Lifetimes presented against $J$ for varying $v$}
\end{figure}

A visualisation of the computed predissociation lifetimes from this work against total angular momentum $J$ is presented in Fig. \ref{fig:ltJ}. For $v>4$ we find that lifetimes behave less predictably, wherein they do not decrease monotonically. The cause of this has not yet been identified with certainty however we expect this to be due to the quasi-bound energy level falling outside of the main crossing region.  

In $v=6$ we see that the lifetime increases and then falls again at $J=15.5$. OH has a high rotational constant of around 20 \cm\   and in the high $J$ limit there is an induced 
rotational barrier. This modifies the potential and introduces a further source of predissociation. In this limit, it would be expected that this added source of predissociation would lower the lifetime. This type of predissociation is particularly dominant in AlH and warrants further study \cite{jt874}.  

\section{Conclusion and Discussion}
\label{sec:conclusion}
A spectroscopic model of OH from \ai potential energy curves and coupling curves, and empirical coupling curves which are refined against latest empirical energies \cite{jt868} is produced using the nuclear motion code \Duo. This spectroscopic model will be extended to
include the \B and \C\ states,  and  combined with \ai dipole moment curves to produce a line list and photoabsorption cross section for hot OH as part of the ExoMol project. This line list will include predissociation lifetimes using the newly extended ExoMol data structure \cite{jt914}; broadening effects due to predissocition have been shown to be important in analyzing astronomical spectra.\cite{jt874}

424 quasi-bound \A\  state rovibronic levels  are processed using the method of Paper I to calculate a set of 374 predissociation lifetimes, $\tau_p$. This method has produced a set of predissociation lifetimes from first principles with uncertainties and provides a satisfactory level of agreement with experimental values. The method used in this paper allows for completeness over all levels considered in a spectroscopic model and is applicable to other systems.  

According with the new ExoMol standards \cite{jt898}, predissociation lifetimes of molecules are to be included into the ExoMol line lists by convolution with the radiative lifetimes  $\tau_r$ as in Eq.~\eqref{eq:tau_tot}. 

The non-zero contribution of the continuum \X\  state levels to the line broadening of the \A\  state levels has been reported and shows that a predissociation into the continuum associated with the ground electronic state is of a small but non-zero probability. In this case, this contribution is negligibly small, however this may not be the case in all systems and warrants consideration in other predissociation studies.
OH exhibits strong and complex predissociation dynamics through the interaction of three repulsive electronic states. It is, however, a streamlined example of this effect as the energetic limits of the repulsive states are the same as that of the ground state. This means that if one ignores asymptotic fine structure effects, the dissociation products are independent of the dissociation channel. 
As this method is generally applicable, molecules with more varied predissociation channels may be of interest for future branching ratio and lifetime studies.  

\section*{Supplementary Material }
The calculated predissociation lifetimes, a comparison with literature data, and the \Duo\ input file are provided as supplementary material,

\section*{Acknowledgements}
This work was supported by the European Research Council (ERC) under the European Union’s Horizon 2020 research and innovation programme through Advance Grant number 883830 and the UK STFC under grant ST/R000476/1. The authors acknowledge the use of the UCL Myriad High Performance Computing Facility (Myriad\@ UCL), and associated support services, in the completion of this work


\section*{Data Availability}
All data is included in the main manuscript or as Supplementary Material. The full lifetime calculation code, \binslt, is freely available at \href{https://github.com/exomol}{ExoMol GitHub}. 


\section*{References}
\bibliography{bibs/journals_phys,
    bibs/jtj,
    bibs/methods,
    bibs/OH,
    bibs/partition,
    bibs/programs,
    bibs/OH_paper,
    bibs/databases}
\end{document}